\documentstyle[prl,aps,preprint]{revtex}

\def\II{\relax{\rm I\kern-.18em I}}
\def\IH{\relax{\rm I\kern-.18em H}}
\def\g#1{\rlap/#1}

\def\frac#1#2{{#1\over#2}}
\def\bold#1{{\bf #1}}
\def\a{(|a|^2+|b|^2)}
\def\b{(|a|^2-|b|^2)}
\def\c{({a^*b}+{ab^*})}

\def\vqs{(v_q^2+a_q^2)}
\def\vqm{(v_q^2-a_q^2)}
\def\vam{(a_q^2-v_q^2)}
\def\ves{(v_e^2+a_e^2)}

\def\ca{\left[\a (P_p\cdot P) +\b m_p M_\Lambda\right]}

\def\zb{{\bar z}}

\def\ie{{\it i.e.\/}}

\def\hff#1{{\hat f}_1^q(#1)}
\def\hgg#1{{\hat g}_1^q(#1)}
\def\hhh#1{{\hat h}_1^q(#1)}
\def\dss{\left(\frac{d\sigma}{d\Omega_q}\right)}

\def\dse{\frac{N_c\alpha^2_{em}}{8s}}
\def\qs{Q_q^2}
\def\cs{(1+\cos^2\Theta)}
\def\ss{\sin^2\Theta}
\def\qx{Q_q\chi_1}
\def\half{\frac{1}{2}}
\begin{document}
\tighten

\title{PROBING QUARK FRAGMENTATION FUNCTIONS \\
FOR SPIN-1/2 BARYON PRODUCTION \\
IN UNPOLARIZED $e^+e^-$ ANNIHILATION\thanks
{This work is supported in part by funds provided by the U.S.
Department of Energy (D.O.E.) under cooperative
agreement \#DF-FC02-94ER40818 and \#DE-FG02-92ER40702.}}

\author{Kun Chen, Gary R. Goldstein\thanks{Permanent address:
Department of Physics, Tufts University, Medford, MA02155.},
R. L. Jaffe, and Xiangdong Ji}

\address{Center for Theoretical Physics \\
Laboratory for Nuclear Science \\
and Department of Physics \\
Massachusetts Institute of Technology \\
Cambridge, Massachusetts 02139 \\
{~}}

\date{MIT-CTP: \#2365  \hfill  HEP-PH: \#9410337  \hfill
 Submitted to: {\it Nucl. Phys. B}
\hfill   September 1994}

\maketitle

\begin{abstract}
We study the measurement of the quark fragmentation functions for
spin-1/2 baryon production ($\Lambda$ and $\bar \Lambda$, in
particular) in unpolarized $e^+e^-$ annihilation.  The spin-dependent
fragmentation functions $\hat g_1(z)$ and $\hat h_1(z)$ can be probed
in the process as a result of quark-antiquark spin correlation and the
weak decay of the baryons.  The relevant cross section is expressed as
a product of the two-jet cross-section, the fragmentation functions,
and the differential width of the hyperon decay.
\end{abstract}

\pacs{xxxxxx}

\section{Introdution}

Due to confinement,
the basic building blocks of QCD, quarks and gluons, cannot
emerge as asymptotic states of the theory. Rather
the high energy quarks and gluons in a hard scattering
process show up in a detector as jets of hadrons,
and their axes
register the momentum directions of the
initial partons. The process
of converting a colored parton into a shower of hadrons
is called parton fragmentation, in which it has long been
known that the soft QCD physics
dominates. As a result, analyzing fragmentation directly
from  first principles is very difficult.
Up to now the fragmentation process is mainly modelled
with a set of empirical rules
and a large Monte Carlo code \cite{AND}. Given the paucity
of our knowledge, experimental information on various
aspects of fragmentation is quite valuable for
understanding the essential physics in the
underlying process.

Semi-inclusive information about fragmentation
is contained in fragmentation
functions, which measure the probability for
a quark to fragment to a specific hadron with
a fixed longitudinal momentum. All the fragmentation
functions experimentally measured so far are spin-independent.
In Refs. \cite{JAF,JI,COLLINS},
the novel spin-dependent fragmentation functions $\hat g_1(z)$
and $\hat h_1(z)$ are
introduced for spin-1/2 baryon production.
Both $\hat g_1(z)$
and $\hat h_1(z)$ are of leading twist and can
produce significant effects in high-energy processes.
The physics of these fragmentation functions is quite simple.
The $\hat g_1(z)$ fragmentation function represents
the probability density to find a spin-1/2 baryon with its
polarization in the direction of the longitudinal polarization of
the original quark.
Likewise,  $\hat h_1(z)$ represents the probability
density to find a spin-1/2  baryon with its polarization in the
direction of the transverse polarization of
the original quark.
Here, longitudinal or transverse is defined relative to the
momentum of the quark.  Undoubtedly, experimental data on these
fragmentation functions contain important information about
spin-transfer during quark fragmentation.

To measure the spin-dependent fragmentation functions,
one normally needs to produce a polarized quark jet.
This is possible with polarized $e^+e^-$ collisions
or polarized deep-inelastic scattering.
On the other hand, in an
unpolarized $e^+e^-$ annihilation, the polarizations
of the created quark and antiquark pair are correlated
due to the chiral invariance of the
electroweak interactions. Therefore if
one can find a way to measure the spin correlation of the
produced hadrons, one can obtain information on the
spin-dependent fragmentation functions. This
observation
forms the foundation for our work.

In this paper, we study the measurement of
the spin-dependent fragmentation functions in
unpolarized $e^+e^-$ annihilation. For definiteness,
we consider correlated production of $\Lambda$
and $\bar \Lambda$ in quark and antiquark jets respectively.
The $\Lambda$ and $\bar\Lambda$ are observed by their weak
decay to nucleon-pion and anti-nucleon-pion
respectively. The weak decay allows one to measure the complete
spin-density matrix of the parent baryon. The spin-dependent
information is extracted from the angular correlation of the
decay products of $\Lambda$ and $\bar \Lambda$.
Our result, of course, is valid for production of
any spin-1/2 baryon which decays weakly to another
spin-1/2 baryon and a pion.

This paper is organized as follows. In \S II, we write down
the differential cross section as a product of the density
matrices for the jet production, the quark fragmentation, and
the baryon decay. We then present the
helicity formalism on which the subsequent calculation
is based. In \S III, we evaluate the
jet production density matrix.
In \S IV, the quark fragmentation density
matrix is defined and expressed in terms of the fragmentation
functions $\hat f_1(z)$, $\hat g_1(z)$, and $\hat h_1(z)$.
In \S V, $\Lambda$ and $\bar \Lambda$
decay density matrices are constructed. And finally in
\S VI, we put everything together to form the
final cross section, and discuss its physical significance.
We summarize our results in
\S VII.  The reader who is interested primarily in our results and
their physical interpretation might omit \S III,\S IV, and \S V and skip
directly to the formulas and discussions in \S VI and \S VII.

 \section{spin Density-matrix formalism in helicity basis}

The calculation of the cross section for $e^-e^+\to q\bar q\to\Lambda
X\bar\Lambda\bar X\to p\pi^-X\bar p\pi^+\bar X$ can be carried out by
following the standard procedure of the operator production expansion,
or equivalently, by using the collinear expansion and ``cut
diagram'' technique \cite{MUTA,MUELLER}. However, since we are interested
in a result at the leading twist, there is a more physical approach
which employs the language of the parton model and follows
the process step by step.
First, we calculate the quark-antiquark production;
then, we deal with quark and antiquark fragmentations;
next, we include the $\Lambda$
and $\bar\Lambda$ decay; and finally, we assemble the $q\bar q$
production, fragmentation and decay processes together to obtain the
experimentally observable
particle distributions.
The only subtlety in the case of a spin-dependent calculation is
that all the intermediate quantities must be in a form
of spin-density matrices.
Then the cross section is expressed as a trace of
product of spin-density matrices for separate subprocesses.
In our case we have,
\begin{eqnarray}
\frac{d^8 \sigma}{d\Omega_{h}dzd\bar z\,
d^2P^{\bot}_{p}\,d^2P^{\bot}_{\bar p}}
           \hskip-4em && \nonumber \\
&\hskip-4em=& \left(\frac{d\sigma(e^-e^+\to
           q\bar q)}{d\Omega_{q}}\right)_{hh',\bar h
           \bar h'}\left(\frac{d\bold{\hat
           M}}{dz}\right)^{h'h}_{HH'}\left(\frac{d\bold{\hat{
           \bar M}}}{{d\bar
           z}}\right)^{\bar h'\bar h}_{\bar H\bar H'}
           \left(\frac{d^2\bold{D}}{d^2P^{\bot}_{p}}\right)^{H'H}
           \left(\frac{d^2\bold{\bar D}}{d^2P^{\bot}_{\bar
           p}}\right)^{\bar H'      \bar H}\ ,
\label{tot1}
\end{eqnarray}
where $h(\bar h)$ and $H(\bar H)$ are indices labelling
spin states of quark (antiquark) and hyperon (antihyperon),
respectively.  By convention, repeated indices are summed over.
The bulk of this paper is devoted to defining and calculating
these spin-density matrices.

One can choose any spin-density matrix formalism to perform
the calculation. The simplest is perhaps the one in which
all the spin indices are just the ordinary Dirac indices.
However, the drawback of this approach is that all the density
matrices are $4\times 4$ and they do not have a clear physical
interpretation.
As is often the case, the physics is much clearer in the helicity
basis\cite{JW}.
Throughout this paper, we shall
use this formalism, and thus $h(h')$ and $H(H')$ in Eq.~(\ref{tot1}) are to be
interpreted as helicity indices with values $\pm 1/2$ (or $\pm$ for
short).

In the helicity formalism, one first choose the helicity basis,
$u(h)$ for fermion and $v(\bar h)$ for antifermion. Then
a general polarization state for fermion (antifermion) can
be expressed as,
\begin{equation}
     U = au(+)+bu(-);\ \ \ V = cv(+) + dv(-)
\end{equation}
or simply in terms of the two-component spinors,
\begin{equation}
     U = \pmatrix{a \cr b}; \quad V = \pmatrix{c \cr
d}. \label{two1}
\end{equation}
Any processes and subprocesses with fermions and/or antifermions
as external particles can be calculated as a density matrix
with pairs of helicity indices. Each pair corresponds to
one particle, with one index representing the spin state of
the particle in the amplitude and the other in the conjugate
amplitude. Obviously, the spin-dependent probability
or cross section can be obtained by contracting these indices
with the two-component wave functions in Eq.~(\ref{two1}).

The helicity formalism treats the polarization of
each fermion (antifermion) independently, since
the two-component wave function is expanded in basis states
which are different for each particle. In particular, the spin
quantization axis is the direction of the particle's momentum.
This is rather convenient because one can quickly write down a
two-component wave function once one knows the relative orientation
between the particle's momentum and its spin vector. In fact,
it will be useful for us
to define a coordinate system for each
particle with its momentum as the $z$-axis.
The directions of
the $\hat x$ and $\hat y$ axes are defined as
the directions of the spin vector $S^\mu=\bar u \gamma^\mu\gamma_5 u$
associated with the following spinors,
\begin{eqnarray}
 u(\hat x) &= &{1\over\sqrt 2}(u(+) + u(-)) \ , \nonumber \\
 u(\hat y) &= &{1\over\sqrt 2}(u(+) + iu(-)) \ .
\end{eqnarray}
This way one is assured that the $\hat x$ and $\hat y$ axes are
defined in the same way for all particles and antiparticles.
The choice of helicity basis for each particle
fixes the relative orientation of the different coordinate systems.
For fermions, the
two-component spin wave function can be chosen universally as,
\begin{equation}
      \pmatrix{a \cr b}=\pmatrix{\cos{\theta\over 2} \cr
          \sin{\theta\over 2}e^{i\phi}}.
\label{wave1}
\end{equation}
where $\theta$ and $\phi$ are polar and azimuthal angles
of the spin vector relative to the momentum and the $x$-axis.

Another important advantage of the helicity formalism
is that an antifermion can be treated exactly like
a fermion. This should be the case because the definition of
fermion and antifermion is itself arbitrary due to
charge conjugation symmetry. However, in ordinary calculations,
some aspects of the symmetry are not obvious because
antifermions are treated as holes in the
negative energy fermion sea. For instance, the spinor
$v$ is associated with antifermion creation,
whereas $u$ is with fermion annihilation. In other words in a
Feynman diagram the momentum flow for antifermions
is always against the fermion number flow, and the Dirac
algebra follows the latter flow.  In the helicity formalism,
if one were to follow the fermion number flow in ordering
the helicity indices, one would find that all the density matrices
would be expressed naturally in terms of the transpose of
Pauli  matrices. Furthermore, if one were to use the charge
conjugation  relation $v=C\bar u^T$ to define the spinor for the
antifermion,  then the two-component wave function would be the
complex conjugate of that in Eq.~(\ref{wave1}),
\begin{equation}
         \pmatrix{c \cr d}= \pmatrix{\cos{\theta\over 2} \cr
          \sin{\theta\over 2}e^{-i\phi}}.
\label{wave2}
\end{equation}
Note again that this wave function is associated with the
antifermion creation. All these observations point to
a simple scheme for dealing with antifermion
spin calculations: Order the
spin indices in terms of the momentum flow rather than the fermion
number flow. Then the spin-density matrices are all naturally
expressed in terms of Pauli matrices and
the two-component wave functions of antifermions
are exactly the same as those of fermions. The wave function in
Eq.~(\ref{wave2}) now appears in a transposed form and is
naturally associated with an antiparticle being created.
One final note: the $\hat x$ and $\hat y$ axes for the antiparticle
are defined by the spin vector $S^{\mu} = -\bar v\gamma^\mu\gamma_5 v$
associated with the spinors,
\begin{eqnarray}
 v(\hat x) &=& {1\over\sqrt 2}(v(+) + v(-)) \ ,     \nonumber \\
 v(\hat y) &=& {1\over\sqrt 2}(v(+) - iv(-)) \ ,
\label{coor}
\end{eqnarray}
where the minus sign follows from Eq.~(\ref{wave2}).

The following sections provide concrete examples for illustrating
the helicity formalism in detail.

\section{Quark-Antiquark Production Density Matrix}

Our calculation begins with
quark-antiquark production via photons ($\gamma^*$)
and $Z^0$s. Here we work in the rest frame of
$\gamma^*$ or $Z^0$. Event by event, we choose the
$z$ axis in the direction of the quark jet,
so the anti-quark momentum is
in the $-z$ direction. [The identification of the quark jet
can be made through study of jet ensembles. However,
in the present calculation we implicitly assume that the $\Lambda$ is a
fragment of the quark. The possibility of
$\bar q\to \Lambda X$ can be included
in the final result in a straightforward way, but the $\Lambda$'s
resulting from $\bar q \to \Lambda X$ will dominantly occur at low
$z$ (the longitudinal momentum fraction of the quark carried by $\Lambda$)
and usually will not satisfy jet isolation cuts. Henceforth we ignore
the possibility of $\bar q \to \Lambda X$ fragmentation.]
By convention, the $y$ axis is defined
by $\hat y=\hat k\times\hat z$, where $\hat k$ is the direction of
the electron beam. In this coordinate system, the 4-momenta
of electron and positron are
\begin{equation}
k^\mu=(E, -E\sin\Theta, 0, E\cos\Theta)\ ,\qquad
k'{}^\mu=(E, E\sin\Theta, 0, -E\cos\Theta),
\label{emom}
\end{equation}
respectively, where $\Theta$ is the polar angle of the electron.

In this section, we seek an expression for the quark-antiquark
production density matrix $\left(d\sigma(e^-e^+\to q\bar q)/
{d\Omega_q}\right)_{hh',\bar h\bar h'}$ in the helicity basis,
where the ordering of the indices follows the
momentum flow of quark and antiquark, and is shown
explicitly in Fig. 1.
While the diagonal elements of the density
matrix are unique, the off-diagonal ones depend on
the phase convention for the helicity basis. In this paper,
we adopt the quark helicity states and the $\gamma$-matrix
representation of Bjorken and Drell \cite{BD}.
The anti-quark helicity states are chosen to be $v(h,k) =
C\bar u(h,k)^T$, where $C=i\gamma^2\gamma^0$ is
the charge conjugation matrix (this differs from \cite{BD} convention).
And thus, in the zero-mass limit, we have,
\begin{equation}
u_q(+)=\sqrt{E}\pmatrix{1\cr 0\cr 1\cr 0}\ ,\quad
u_q(-)=\sqrt{E}\pmatrix{0\cr 1\cr 0\cr -1}\ , \quad
v_{\bar q}(+)=\sqrt{E}\pmatrix{1\cr 0\cr -1\cr 0}\ ,\quad
v_{\bar q}(-)=\sqrt{E}\pmatrix{0\cr 1\cr 0\cr 1}
\end{equation}
Once again, the antiquark is moving in the $-z$
direction.

Clearly the coordinate system for treating quark polarization is
just the one we have defined. On the other hand,
the coordinate system associated with
the anti-quark polarization ($\bar x, \bar y, \bar z$)
can be worked out from our
choice for the antiquark helicity states and the
definition for the axes in the last section.
The $\bar z$ axis is clearly in the
$-z$ direction. Using Eq.~(\ref{coor}),
we find that the $\bar x$ axis is
in the direction of the $x$ axis
and the $y$ axis in the
$-y$ direction. [To avoid a kinematic zero,
one has to restore the quark mass
in the helicity states.]
This coordinate system will be
used for anti-particles throughout this paper.

A straightforward calculation with the standard-model electroweak
currents and the above helicity states yields,
\begin{eqnarray}
\dss_{++,--}&\!\!=&\dse\biggl\{\qs\cs+\chi_2(v_q-a_q)^2\left[\cs\ves-4v_ea_e
           \cos\Theta\right] \nonumber \\ &&
\qquad -2\qx(v_q-a_q)\left[v_e\cs-2a_e\cos\Theta\right]\biggr\}\ ,
           \nonumber \\
\dss_{--,++}&\!\!=&\dse\biggl\{\qs\cs+\chi_2(v_q+a_q)^2\left[\cs\ves+4v_ea_e
           \cos\Theta\right] \nonumber \\ &&
\qquad -2\qx(v_q+a_q)\left[v_e\cs+2a_e\cos\Theta\right]\biggr\}\ ,
           \nonumber  \\
\dss_{+-,-+}&\!\!=&\dse\ss\Biggl\{\qs-\chi_2\ves\vam-2\qx
           v_e\biggl(v_q+\frac{i\Gamma_ZM_Z}{s-M^2_Z}a_q\biggr)\Biggr\}\ ,
	\nonumber \\
\dss_{-+,+-}&\!\!=&\dse\ss\Biggl\{\qs-\chi_2\ves\vam-2\qx
           v_e\biggl(v_q-\frac{i\Gamma_ZM_Z}{s-M^2_Z}a_q\biggr)\Biggr\}\ ,
	\nonumber \\
\dss_{\rm others}\>&\!\!=&0 \ ,
\label{jetden1}
\end{eqnarray}
where $v_e=4\sin^2\theta_W-1$ and $a_e=-1$ are the vector and axial
vector couplings of the electron to the $Z$.  The couplings of the
quarks to the $Z$ are $v_u=1-\frac{8}{3}\sin^2\theta_W$, $v_d=v_s=-1
+\frac{4}{3}\sin^2\theta_W$, $a_u=1$, and $a_d=a_s=-1$.
$N_c=3$ is the color number of the quark, and
\begin{eqnarray}
\chi_1&=&{1\over 16\sin^2\theta_W\cos^2\theta_W} {s(s-M_Z^2) \over
           (s-M_Z^2)^2+\Gamma_Z^2M_Z^2}\ , \\
\chi_2&=&{1\over 256\sin^4\theta_W\cos^4\theta_W}{s^2 \over (s-M_Z^2)^2
           +\Gamma_Z^2M_Z^2}\ .
\end{eqnarray}
Eq.~(\ref{jetden1}) indicates that the quark and antiquark from the
same vertex have opposite helicity, a consequence of the
massless limit. Therefore, as we will see in the next section,
the polarizations of $\Lambda$ and $\bar\Lambda$ are correlated
through the effects of spin-dependent quark fragmentation
functions.

The cross-section density matrix
can also be expressed in terms of the Pauli matrices,
\begin{eqnarray}
\dss&=&\dse\Biggl[\biggl\{\half\qs\cs  \nonumber \\
&&\qquad\qquad\qquad
   +\half\chi_2\Bigl[\cs\vqs\ves
   +8v_ea_ev_qa_q\cos\Theta\Bigr] \nonumber \\
&&\qquad\qquad\qquad -\qx\Bigl[v_ev_q\cs+2a_ea_q\cos\Theta\Bigr]\biggr\}
   (\II^q\otimes\II^{\bar q}-
   \sigma_z^q\otimes \sigma_{\bar z}^{\bar q}) \nonumber\\
&&\qquad+\biggl\{\chi_2\Bigl[v_qa_q\ves\cs+2v_ea_e\vqs\cos\Theta
   \Bigr] \label{res1} \\&&
   \qquad\qquad\qquad-\qx\Bigl[a_qv_e\cs+2v_qa_e\cos\Theta\Bigr]\biggr\}
   (\II^q\otimes\sigma_{\bar z}^{\bar q}
   -\sigma_z^q\otimes\II^{\bar q}) \nonumber \\
&&\qquad+\biggl\{\half\qs+\half\chi_2\ves\vqm-\qx v_ev_q\biggr\}\ss
   (\sigma_x^q\otimes \sigma_{\bar x}^{\bar q} + \sigma_{y}^q\otimes
   \sigma_{\bar y}^{\bar q}) \nonumber \\&
&\qquad-\qx v_ea_q\frac{\Gamma_ZM_Z}{s-M_Z^2}\ss
   (\sigma_x^q\otimes \sigma_{\bar y}^{\bar q}-\sigma^{q}_y\otimes
   \sigma_{\bar x}^{\bar q})\Biggr]\ . \nonumber
\end{eqnarray}

Any physical cross section can be
obtained by taking the trace of the cross
section density matrix with quark-antiquark polarization
density matrices, which are defined as
\begin{equation}
      D^q = \psi'_q\psi^\dagger_q; \ \ D^{\bar q}=\psi_{\bar q}^{'}
     \psi^{\dagger}_{\bar q}.
\end{equation}
For example, if we want to calculate the cross
section when both quark and antiquark are polarized along the $x$
axis, we have,
\begin{equation}
\frac{d\sigma(e^-e^+\to q(\hat
x)\bar q(\hat x))}{d\Omega_q}
=\hbox{tr}\left[\left(\frac{d\sigma}{d\Omega_q}\right)(D^q
\otimes D^{\bar q})\right]\ .
\end{equation}
where $D^q=\II^q + \sigma_x^q$ and $D^{\bar q}=\II^{\bar q} +
\sigma_{\bar x}^{\bar q}$ are the density matrices
for quark and antiquark polarization, respectively.

\section{Fragmentation Density Matrix Formalism}

Quark fragmentation functions were introduced to describe
hadron production from the underlying hard-parton processes.  Apart from the
well-known, spin-independent and chiral-even fragmentation function
${\hat f}_{1}(z)$, there exist various chiral-odd and spin-dependent
fragmentation functions which are of particular interest
because they describe novel spin effects in
hadron production\cite{JAF,JI}. At the leading twist,
there are two additional fragmentation functions for
spin-1/2 baryon production, $\hat g_1(z)$ and $\hat h_1(z)$.
All these fragmentation functions can be expressed
as light-cone correlations in QCD,
\begin{equation}
{\hat f}_{1}(z)=\frac{1}{4}z\int\frac{d\lambda}{2\pi}e^{-i\lambda/z}
\Bigl\langle 0|{\g n}\psi(0)|\Lambda(PS)X\Bigr\rangle
\left\langle\Lambda(PS)X|
{\bar\psi}(\lambda n)|0\right\rangle\ ,
\label{f1}
\end{equation}
\begin{equation}
{\hat
g}_{1}(z)=\frac{1}{4}z\int\frac{d\lambda}{2\pi}e^{-i\lambda/z}
\left\langle 0|{\g n}\gamma_{5}\psi(0)|\Lambda(PS_{\|})X\right\rangle
\left\langle\Lambda
(PS_{\|})X|{\bar\psi}(\lambda n)|0\right\rangle\ ,
\label{g1}
\end{equation}
\begin{equation}
{\hat h}_{1}(z)=\frac{1}{4}z\int\frac{d\lambda}{2\pi}e^{-i\lambda/z}
\Bigl\langle0|{\g n}\gamma_{5}{\g S}_{\bot}\psi(0)|\Lambda(PS_{\bot})X
\Bigr\rangle
\left\langle\Lambda(PS_{\bot})X|{\bar\psi}(\lambda n)|0\right\rangle\ ,
\label{h1}
\end{equation}
where $\Lambda(PS)$ represents the $\Lambda$ hyperon with
four-momentum $P^{\mu}$ and polarization $S^{\mu}$,
normalized to $P^2 = M_\Lambda^2$ and $S^2 = - M_\Lambda^2$
respectively.  We write
$P^{\mu}=p^{\mu}+n^{\mu}M_{\Lambda}^{2}/2$,
$S^{\mu}= S^\mu_\| +M_{\Lambda}S_{\bot}^{\mu}$,
and $S^{\mu}_\|=S\cdot np^{\mu}+S\cdot pn^{\mu}$
with $p^{\mu}$ and $n^{\mu}$ two null vectors ($p^{2}=n^{2}=0$
$p\cdot n = 1$). The special component of
$p$ ($n$) is along (opposite to) the direction of
the $\Lambda$ momentum. These light-cone vectors will be
defined separately for every observable hadron and will be
labelled accordingly.
The variable $z$, representing the
momentum fraction of the quark carried by $\Lambda$, is defined
in the usual way
$z=2P\cdot q/q^2$, where $q$ is the momentum carried by the virtual
boson. The
summation over $X$ is implicit and covers all possible states which
can be populated by the quark fragmentation, and also the
renormalization point ($\mu^{2}$) dependence is suppressed.  [QCD
radiative corrections generate a different $\mu^2$ dependence for each
moment of these fragmentation functions, which is associated with the
$Q^2$ evolution of the experimental data.  Although important for
comparison with experiment, QCD evolution does not disturb the
classification of spin dependent effects, so we suppress it throughout
this analysis.]

We turn to the density matrix formalism for
quark fragmentation functions. Here we aim to
construct quark and antiquark fragmentation density matrices that
depend on helicity indices of quark (antiquark) and
hyperon (antihyperon). Such a construction can proceed
in three steps. First, we construct a density matrix
in the Dirac representation.
Then, we obtain a density matrix in
the mixed representation, in which the helicity indices of
hyperon (antihyperon) are manifest, but the quark
(antiquark) is still labelled by the Dirac indices.
In the final step, we transform the
remaining Dirac indices into helicity indices.

For the quark fragmentation $q\rightarrow\Lambda X$, we define
\begin{equation}
{\hat M}_{\Lambda}(z, S, P)_{\alpha\beta}=
\ \int\frac{d\lambda}{2\pi}e^{-i\lambda/z}
\Bigl\langle 0|\psi_{\alpha}(0) | \Lambda(PS)X\Bigr\rangle
\Bigl\langle\Lambda(PS)X|{\bar\psi}_{\beta}(\lambda n) | 0\Bigr\rangle\ \ .
\label{qden1}
\end{equation}
Then, combining with Eqs.~(\ref{f1} -- \ref{h1}), we have,
\begin{equation}
{\hat M}_{\Lambda}(z, S, p/z)=\frac{{\hat f}_{1}(z)}{z}{\g p}
+\frac{{\hat g}_{1}(z)}{z}(n\cdot S_{\|})\gamma_{5}{\g p} +\frac{{\hat
h}_{1}(z)}{z}\gamma_{5}{\g S}_{\bot}{\g p}\ \ .
\label{qden2}
\end{equation}
to the leading twist. Another set of fragmentation
functions can be defined to describe anti-quark
fragmentation ${\bar q}\rightarrow {\bar\Lambda}\, {\bar X}$.
However, they can be related to the above
fragmentation functions by using charge conjugation. In fact, if
we define the
anti-quark fragmentation density matrix as
\begin{equation}
{\hat{\bar M}}_{\bar\Lambda}(\bar z, \bar S, \bar P)_{\alpha\beta} =
\int\frac{d\lambda}{2\pi}e^{-i\lambda/\bar z}
\Bigl\langle 0| {\bar\psi}_{\beta}(0)|{\bar\Lambda}(\bar P\bar S)\bar
X\Bigr\rangle
\Bigl\langle {\bar \Lambda}(\bar P \bar S)\bar X
|\psi_{\alpha}(\lambda \bar n)|0\Bigr\rangle
\ \ ,
\label{qbden1}
\end{equation}
we find by using charge conjugation,
\begin{equation}
{\hat{\bar M}}_{\bar\Lambda}(\bar z, \bar S, \bar p/\bar z)=-C^{-1}{\hat M}
_{\Lambda}^{T}(\bar z, \bar S, \bar p/\bar z)C\ \ ,
\label{qbden2}
\end{equation}
where ${\hat M}_{\Lambda}^{T}(z, S, p/z)$ means the transpose of matrix
${\hat M} _{\Lambda}(z, S, p/z)$.
Consequently for the anti-quark fragmentation
$\bar q\rightarrow \bar\Lambda\bar X$,
\begin{equation}
{\hat{\bar M}}_{{\bar\Lambda}}(\bar z, \bar S, \bar p/\bar z)
=\frac{{\hat f}_{1}(\bar z)}
{\bar z}{\g {\bar p}}-\frac{{\hat g}_{1}(\bar z)}{\bar z}(\bar
n\cdot \bar S_{\|})\gamma_{5} {\g {\bar p}}
+\frac{{\hat h}_{1}(\bar z)}{\bar z}\gamma_{5} {\g {\bar S}}_{\bot}
{\g {\bar p}}\ \ .
\label{qbden3}
\end{equation}
Notice the sign change for the ${\hat g}_{1}(\bar z)$ term.

The next step is to change $\hat M_\Lambda$, which
depend functionally on the spin $S^\mu$ of the $\Lambda$, to a $2
\times 2$ fragmentation density matrix carrying
$\Lambda$ ($\bar \Lambda$) helicity indices. The density matrix is
defined as,
\begin{equation}
{\hat M}(z, P)_{\alpha\beta}^{HH'} =
\ \int\frac{d\lambda}{2\pi}e^{-i\lambda/z}
\Bigl\langle 0|\psi_{\alpha}(0)
|\Lambda(PH')X\Bigr\rangle
\Bigl\langle\Lambda(PH)X|{\bar\psi}_{\beta}(\lambda n)
|0\Bigr\rangle\ \ .
\label{qden3}
\end{equation}
The transformation is possible because Eq.~(\ref{qden2}) with a general
$S$ contains all the information about the
$\Lambda$ spin dependence of the quark
fragmentation. If we choose $S$
to reproduce the helicity eigenstates,
the diagonal elements of the spin density matrix are immediately obtained,
\begin{eqnarray}
{\hat M}_{++}=\   {\hat M}_{\Lambda}(z, S^\mu=
{e}_{z}^{\mu},
 p/z)&=&\frac{{\hat f}_{1}(z)}{z}{\g p}+\frac{{\hat g}_{1}(z)}{z}
\gamma_{5}{\g p}, \nonumber \\
{\hat M}_{--}=\  {\hat M}_{\Lambda}(z, S^\mu=
-{e}_z^{\mu},
 p/z)&=&\frac{{\hat f}_{1}(z)}{z}{\g p}-\frac{{\hat g}_{1}(z)}{z}
\gamma_{5}{\g p},
\label{hel1}
\end{eqnarray}
where we have chosen the direction of the $\Lambda$ momentum as the
$z$ direction. [In the leading twist calculation, the direction
of the quark and $\Lambda$ can be taken to be collinear. The contributions
from the $\Lambda$ transverse momentum are among higher twists.]
According to the superposition principle, we may
extract the off-diagonal elements of the
spin density matrix from Eq.~(\ref{qden2}). From
\begin{equation}
|S^\mu=M_\Lambda e_x^\mu\rangle=\frac{1}{\sqrt{2}}\pmatrix{1\cr 1},
\hskip .5truein |S^\mu = M_\Lambda e_y^\mu\rangle=
\frac{1}{\sqrt{2}}\pmatrix{1\cr i},
\end{equation}
we have
\begin{equation}
{\hat M}_{\Lambda}(z,S^\mu=M_\Lambda e_x^\mu
, p/z)=\frac{1}{2}
\bigl({\hat M}_{++}+{\hat M}_{+-}+{\hat M}_{-+}+{\hat M}_{--}
\bigr)\ \ ,
\end{equation}
\begin{equation}
{\hat M}_{\Lambda}(z,S^\mu=M_\Lambda e_y^\mu, p/z)=\frac{1}{2}
\bigl({\hat M}_{++}+i{\hat M}_{+-}-i{\hat M}_{-+}+{\hat M}_{--}
\bigr)\ \ .
\end{equation}
{}From these relations we can read off the off-diagonal
components of $\hat M$ in the
$\Lambda$ helicity basis,
\begin{eqnarray}
{\hat M}_{+-}&=&
\frac{{\hat h}_{1}(z)}{z}\gamma_{5}({\g e}_{x}-i
{\g e}_{y}){\g p}, \label{hel2} \\
{\hat M}_{-+}
&=&\frac{{\hat h}_{1}(z)}{z}\gamma_{5}({\g e}_{x}+i
{\g e}_{y}){\g p}. \nonumber
\end{eqnarray}
Here, the four-vectors $e^{\mu}_x$, $e^{\mu}_y$ and $e^\mu_z$ are
defined by
$e^\mu_x = (0,1,0,0)$, $e^\mu_y = (0,0,1,0)$ and
$e^\mu_z=(P^z,0,0,P^0)$. In terms of the Pauli
matrices $\{\sigma^{\Lambda}_k\}$ and $2\times 2$ identity matrix
$\II{^\Lambda}$, Eqs.~(\ref{hel1}) and (\ref{hel2})
can be summarized as,
\begin{equation}
{\hat M}_{\Lambda}(z,p/z)=\frac{{\hat f}_{1}(z)}{z}{\g p}
\ \II{^\Lambda}+\frac{{\hat g}_{1}(z)}{z}\gamma_{5}{\g
p} \ \sigma^{\Lambda}_z-\frac{{\hat h}_{1}(z)}{z}
\gamma_{5}{\g p}
\left( {\g e}_{x}\sigma^{\Lambda}_x+{\g
e}_{y}\sigma^{\Lambda}_y \right).
\end{equation}

For the antiquark fragmentation, we define
the density matrix analogous to Eq.~(\ref{qden3}),
\begin{equation}
{\hat {\bar M}}(\bar z, \bar P)_{\alpha\beta}^{\bar H\bar H'} =
\ \int\frac{d\lambda}{2\pi}e^{-i\lambda/\bar z}
\Bigl\langle 0|\bar \psi_{\beta}(0)
|\bar \Lambda(\bar P\bar H')\bar X\Bigr\rangle
\Bigl\langle \bar \Lambda(\bar P\bar H)\bar X|{\psi}_{\alpha}(\lambda \bar n)
|0\Bigr\rangle\ \ .
\label{qden31}
\end{equation}
Then a totally parallel analysis shows that
\begin{equation}
{\hat{\bar M}}_{\bar\Lambda}(\zb,\bar p/\zb)
=\frac{{\hat f}_{1}(\zb)}{\zb}
{\g {\bar p}}\ \II^{\bar\Lambda}-\frac{{\hat g}_{1}(\zb)}
{\zb}\gamma_{5}{\g {\bar  p}}\ \sigma^{\bar\Lambda}_{\bar z} -
\frac{{\hat h}_{1}(\zb)}{\zb}
\gamma_{5}{\g{\bar  p}}
\left( {\g e}_{\bar x}\sigma^{\bar\Lambda}_{\bar x}+
{\g e}_{\bar y}\sigma^{\bar\Lambda}_{\bar y} \right)\ ,
\end{equation}
with similar definitions
\begin{eqnarray}
{\hat{\bar M}}_{++}&=&{\hat{\bar
M}}_{\bar\Lambda}(\zb, \bar S^\mu=e_{\bar z}^\mu, \bar p/\zb)\ , \nonumber \\
{\hat{\bar M}}_{--}&=&{\hat{\bar
M}}_{\bar\Lambda}(\zb,\bar S^\mu=-e_{\bar z}^\mu, \bar p/\zb)\ .
\end{eqnarray}
Here, the four-vectors $e_{\bar x}^{\mu}$, $e_{\bar y}^{\mu}$
and $e_{\bar z}^\mu$
are defined by $e_{\bar x}^\mu = (0,1,0,0)$, $e_{\bar
y}^\mu = (0,0,1,0)$ and
$e_{\bar z}^\mu=(\bar P^{\bar z},0,0,\bar P^0)$ in the frame where
the $\bar\Lambda$ momentum is in the $\bar z$-direction.

Finally we calculate the density matrices entirely in
the helicity basis. This reformulation is straightforward---the Dirac
matrix form of ${\hat M}_{\Lambda}(z,p/z)$ is inserted between
the helicity basis states $\bar{u}(h')$ and $u(h)$.
After taking care of the proper normalization,
the resulting $2\times 2$ matrix, whose elements are labelled by the
quark helicities, $h'h$, can be expanded
in a quark Pauli matrix basis, as was just done for the $\Lambda$
indices. The result has the remarkably simple direct product form,
\begin{equation}
\frac{d{\hat {\bold{M}}}_{\Lambda}}{dz}(z,p/z)={\hat f}_{1}(z)
\ \II^{q}\otimes \II^\Lambda+{\hat g}_{1}(z) \ \sigma_z^q
\otimes \sigma_{z}^\Lambda+{\hat h}_{1}(z)
\left( \sigma_x^q\otimes \sigma_{x}^\Lambda+\sigma_{y}^q\otimes
\sigma_{y}^\Lambda \right) .
\label{pro1}
\end{equation}
It is clear from this form that ${\hat g}_1$ measures the probability
that the longitudinal polarization of the quark transferred to that of
$\Lambda$ with fractional momentum $z$. The ${\hat h}_1$ measures the
probability that the transversity of the quark is transferred.
There is an obvious invariance
of the ${\hat h}_1$ term under rotations about the momentum direction,
since transverse momenta have been integrated out of the fragmentation
process. An analogous expression can be obtained for the anti-quark
fragmentation into $\bar{\Lambda}$, where the corresponding Pauli
matrices will be defined in terms of the anti-particle's
helicities. The direct product form for the antiquark case is
\begin{equation}
\frac{d{\hat {\bold{\bar
M}}}_{\bar\Lambda}}{d\zb}(\zb, \bar p/\zb)={\hat f}_{1}(\zb)
\ \II^{\bar q}\otimes \II^{\bar\Lambda}+{\hat g}_{1}(\zb) \
\sigma_{\bar z}^{\bar q}\otimes \sigma^{\bar\Lambda}_{\bar z}
+{\hat h}_{1}(\zb)
\left( \sigma_{\bar x}^{\bar q}\otimes \sigma^{\bar\Lambda}_{\bar x}+
	\sigma^{\bar q}
_{\bar y}\otimes\sigma^{\bar\Lambda}_{\bar y} \right) .
\label{pro2}
\end{equation}
The fact that Eqs.~(\ref{pro1}) and (\ref{pro2}) are identical
is a direct consequence of charge conjugation symmetry,
which is manifest in the formalism developed in \S II.

An expression like Eq.~(\ref{pro1}) could have been written upon
consideration of the rotation invariance and parity conservation of
the parton fragmentation process. Aside from an explicit
derivation, what is novel here is the
identification of the coefficients with the leading-twist
fragmentation functions, which have a
formal connection with QCD.

\section{$\Lambda$ ($\bar\Lambda$) Decay Density Matrix}

It is well-known that the polarization of
$\Lambda$ and $\bar \Lambda$
can be measured through their weak decay.
The relevant formalism and experimental
information about their decay
are included in the standard particle data book.
Here, for our purpose, we need to obtain
the spin density matrices for the decay.

For the hyperon nonleptonic decay,
the most general decay amplitude is \cite{CB}
\begin{equation}
{\cal M}=G_{F}m_{\pi}^{\,\,2}{\bar
u}_{f}(A-B\gamma_{5})u_{i},
\label{amp1}
\end{equation}
where $A$ and $B$ are constants and generally complex numbers, and
$u_i$ and $u_f$ are Dirac spinors for the initial and final baryons,
respectively.  Analogous to Eq.~(\ref{amp1}), the most
general amplitude for the anti-hyperon nonleptonic decay is
\begin{equation}
{\cal M}'=-G_{F}m_{\pi}^{\,\,2}{\bar v}_{i}(A^*+B^*
\gamma_{5})v_{f}
\label{amp2}
\end{equation}
with $v_i$ and $v_f$ Dirac spinors for the initial and final
anti-baryons respectively.
The spin density matrices for the $\Lambda$
and $\bar\Lambda$ decay can be defined as,
\begin{eqnarray}
D_{H'H} & = & \sum_{s_{p}}
\Bigl\langle p\pi^-|\Lambda(PH)\Bigr\rangle
\Bigl\langle\Lambda(P H')|p\pi^{-}\Bigr\rangle\ \ ,  \nonumber  \\
{\bar D}_{\bar H'\bar H} &=& \sum_{s_{\bar p}}
\Bigl\langle \bar p\pi^{+}|\bar\Lambda(\bar P\bar H)\Bigr\rangle
\Bigl\langle\bar \Lambda(\bar P\bar H')|\bar p\pi^{+}\Bigr\rangle\
\ .
\label{dec1}
\end{eqnarray}
We have neglected the phase space integration which
can be discussed separately.
Substituting Eq.~(\ref{amp1}) into Eq.~(\ref{dec1})
and again using the superposition
principle, we find the four elements of
$D_{H'H}$,
\begin{eqnarray}
D_{++}&=&2\left[{\a P_{p}\cdot
P}+{\b m_{p}M_{\Lambda}}+{\c P_{p}\cdot S_{R}}\right]\ \ , \nonumber\\
D_{--}&=&2\left[{\a P_{p}\cdot P}+{\b
m_{p}M_{\Lambda}}-{\c P_{p}\cdot S_{R}}\right]\ \ , \nonumber \\
D_{+-}&=&2\c M_{\Lambda}P_{p}\cdot(e_{x}-ie_{y})\ \ , \\
D_{-+}&=&2\c M_{\Lambda}P_{p}\cdot(e_{x}+ie_{y})\ \ , \nonumber
\end{eqnarray}
 where
$S_{R}$ is related to the light-cone variables through
$S_{R}=p-n\,M_{\Lambda}^{2}/2$, $P_p$ is the momentum
of the proton, and
\begin{eqnarray}
 a&=& G_fm_{\pi^{+}}^{\,\,2}\,A\ , \nonumber \\
 b&=&-G_fm_{\pi^{+}}^{\,\,2}\,B\ .
\end{eqnarray}

In terms of Pauli matrices, we obtain a compact form
for the density matrix,
\begin{eqnarray}
 {\bold D}&=& 2\left[{\a P_{p}\cdot P}+{\b m_{p}M_{\Lambda}}\right]
           \, \II^\Lambda \nonumber \\
 && \quad +2\c (P_{p}\cdot
           S_{R}) \, \sigma^{\Lambda}_z \nonumber \\
 && \quad +2\c M_{\Lambda}\left[(P_{p}\cdot e_x)
           \, \sigma^{\Lambda}_x + (P_{p}\cdot e_y) \, \sigma^{\Lambda}_y
\right]\ \ .
\label{gary1}
\end{eqnarray}
A similar calculation for $\bar\Lambda$ gives
\begin{eqnarray}
\bold{\bar D} &=&  2\left[{\a P_{\bar p}\cdot \bar P}+{\b
           m_{p} M_{\Lambda}}\right]\, \II^{\bar\Lambda} \nonumber \\
&& \quad -2\c(P_{\bar p}\cdot \bar S_{R}) \, \sigma^{\bar\Lambda}_{\bar z}
           \nonumber \\
&& \quad -2\c M_{\Lambda}\left[(P_{\bar p}\cdot e_{\bar x})
           \, \sigma^{\bar\Lambda}_{\bar x}+(P_{\bar p}\cdot e_{\bar y}) \,
           \sigma^{\bar\Lambda}_{\bar y}\right] \> \ ,
\label{gary2}
\end{eqnarray}
where $P_{\bar p}$ is the momentum of the antiproton and
$\bar S_R = \bar p -\bar n M^2_{\Lambda}/2$.

When performing the phase space integration, we keep
the transverse momenta of the proton and the antiproton as
differential variables.  To take into account
the particle decay width and the motion of the parent particle, it is
convenient to define the boost-invariant quantities
\begin{eqnarray}
\frac{d^2\bold{D}}{d^2 P_p^\bot}&=&\frac{1}{2M_\Lambda\Gamma_\Lambda}
\int \frac{dP_p^z}{(2\pi)^3 2E_p}\frac{d^3P_\pi}{(2\pi)^3 2E_\pi}
\ \bold{D}\ (2\pi)^4\delta^{(4)}(P-P_p-P_\pi),\ \nonumber \\
\frac{d^2\bold{\bar D}}{d^2 P_{\bar p}^\bot}&=&\frac{1}{2M_{\Lambda}
\Gamma_{\Lambda}}
\int \frac{dP_ {\bar p}^z}{(2\pi)^3 2E_{\bar p}}\frac{d^3P_\pi}
{(2\pi)^3 2E_\pi}\ \bold{{\bar D}}\ (2\pi)^4\delta^{(4)}
(\bar P-P_{\bar p}-P_\pi),\
\label{dif1}
\end{eqnarray}
where $\Gamma$ is the total width of the $\Lambda$ decay and
$P_\pi$ is the momentum of the pion.
Particle masses have to be kept
explicitly in the decay processes although they can be neglected in
high-energy quark fragmentation.  After integration,
Eq.~(\ref{dif1}) becomes,
\begin{eqnarray}
\frac{d^2\bold{D}}{d^2 P_p^\bot}&=&\frac{\bold{D}}{8(2\pi)^2M_\Lambda
\Gamma_\Lambda\left|
P^{-}P_{p}^{+}-P^{+}P_{p}^{-}\right|}\ , \nonumber \\
\noalign{\smallskip}
\frac{d^2\bold{{\bar D}}}{d^2 P_{\bar
p}^\bot}&=&\frac{\bold{\bar D}}{8(2\pi)^2M_{\Lambda}\Gamma_{\Lambda}
\left|{\bar P}^{-}P_{\bar p}^{+}-{\bar P}^{+}P_{\bar
p}^{-}\right|}\ .
\label{res4}
\end{eqnarray}
where $P^{\pm}$ are defined in the usual light-cone coordinates,
$P^{\pm} = (P^0 \pm P^z)/\sqrt{2}$.

\section{Cross Section for $\bf{\lowercase{e^- e^+ \to q\bar q
\to\Lambda} X \bar\Lambda\bar X\to \lowercase {p}
\pi^-X \lowercase {\bar p}\pi^+\bar X}$ And
Its physical significance}

Now we have the spin density matrices for
all the subprocesses in $e^-e^+\to q\bar
q\to\Lambda X\bar\Lambda\bar X\to p\pi^-X\bar p\pi^+\bar X$.
Substituting Eqs.~(\ref{res1}), (\ref{pro1}), (\ref{pro2}), and
(\ref{res4}) to Eq.~(\ref{tot1}),
we get the final cross section
\begin{eqnarray}
\frac{d^8\sigma}
           {d\Omega_\Lambda dzd\zb\,d^2P_p^\bot\, d^2P_{\bar p}^\bot}
           \hskip-4em && \nonumber \\
&\hskip-4em=& \frac{1}
           {2(2\pi)^2M_\Lambda\Gamma_\Lambda\left|P^-P_p^+-P^+P_p^-\right|}
           \cdot\frac{1}
           {2(2\pi)^2 M_{\Lambda} \Gamma_{\Lambda} \left|{\bar P}^-
           P_{\bar p}^+-{\bar P}^+P_{\bar p}^-\right|}
           \nonumber \\
&\hskip-4em&\times  \frac{N_c\alpha^2_{em}}{4s}\cdot
           \ca^2 , \nonumber \\
&\hskip-4em& \times \sum_{q=u,d,s}
           \Biggl[ \biggl\{ \qs\cs+\chi_2 \Bigl[
           \cs\vqs\ves+8v_ea_ev_qa_q \cos\Theta\Bigr]
           \nonumber \\
&\hskip-4em& \qquad
           -2\qx \Bigl[v_ev_q\cs+2a_ea_q\cos\Theta\Bigr] \biggr\}
           \Bigl( \,\hff{z}\hff{\zb}+C_{gg}\,\hgg{z}\hgg{\zb} \Bigr)
           \nonumber \\
\hskip-4em&\hskip-4em& -\biggl\{\chi_2\Bigl[
           2v_qa_q\ves\cs+4v_ea_e\vqs\cos\Theta\Bigr]
           \nonumber\\
&\hskip-4em&\qquad -2\qx\Bigl[a_qv_e\cs+2v_qa_e\cos\Theta\Bigr]\biggr\}
           \Bigl( C_{fg}\,\hff{z}\hgg{\zb}+C_{gf}\,\hgg{z}\hff{\zb} \Bigr)
           \nonumber \\
&\hskip-4em&-\biggl\{\Bigl[ \qs+\chi_2\ves\vqm-2\qx v_ev_q\Bigr]
           \cos(\varphi+\bar\varphi) \nonumber \\
&\hskip-4em&\qquad
           +2\qx \frac{\Gamma_ZM_Z}{s-M^2_Z}
           v_ea_q\sin(\varphi+\bar\varphi) \biggr\}
           \ss\ C_{hh}\,\hhh{z}\hhh{\zb}\Biggr]\ ,
\label{fin}
\end{eqnarray}
with $\varphi$ and $\bar\varphi$ the azimuthal angles of the proton
and the antiproton, respectively, in the coordinate system
that is defined in the beginning of \S III: the $z$ axis
is chosen to be the direction of the quark jet (or $\Lambda$)
and the $x$ axis in the plane of the beams and jets.
The $C$'s are given by
\begin{eqnarray}
C_{gg}&=&\alpha^2 {(P_p\cdot S_R)(P_{\bar p}\cdot \bar S_R)\over M_\Lambda^2
     m_p^2}=\alpha^2{|P_{p\|}|\cdot
|P_{\bar p\|}| \over m_p^2}{\eta\bar \eta} \ , \nonumber \\
C_{fg}&=&\alpha {(P_{\bar p}\cdot \bar S_R) \over M_\Lambda m_p}=\alpha
{|P_{\bar p\|}| \over m_p}\bar \eta \ , \\
C_{gf}&=&\alpha {(P_{ p}\cdot S_R) \over M_\Lambda m_p} =
\alpha {|P_{p\|}| \over m_p}\eta \ , \nonumber \\
C_{hh}&=&\alpha^2 {\left|{\vec{P_p}}{\!}^\bot\right|
\left|{\vec{P_{\bar p}}}{\!}^\bot\right| \over m_p^2} \ . \nonumber
\end{eqnarray}
where $P_p\cdot
P =(M_\Lambda^2+m_p^2-m_\pi^2)/2$ in $C_{ff}$
and $\eta$ ($\bar \eta$) is $\pm 1$ depending on whether the
momenta of proton (antiproton) and $\Lambda$ ($\bar \Lambda$)
are parallel or antiparallel. $P_{p\|}$ and $P_{\bar p\|}$ are
the projections of the proton and antiproton momenta in
the directions of $\Lambda$ and $\bar \Lambda$, respectively,
in the respective {\it rest frames\/} of the parent particles,
and ${\vec{P_p}}{\!}^\bot$ and
${\vec{P_{\bar p}}}{\!}^\bot$ are projections of the proton
and antiproton momenta onto the $x$-$y$ plane. The $\alpha$
is the standard hyperon-decay parameter
defined in the particle date table \cite{PDG}.

There are three distinct classes of terms in the cross section.
They correspond to three different type of angular dependences and are
sensitive to different combinations of fragmentation functions.
The first class involves terms with products of fragmentation
functions $\hat f_1(z) \hat f_1(\bar z)$
and $\hat g_1(z) \hat g_1(\bar z)$. They have no azimuthal dependence. The
dependence on the polar angle arises entirely from the two-jet
production cross section. To isolated the $\hat g_1(z) \hat g_1(\bar z)$
term, one has to measure
the correlation of the proton and antiproton momenta with
respect to the momenta of $\Lambda$ and $\bar \Lambda$.
The second class of terms contains the products of fragmentation
functions $\hat f_1(z)$ and $\hat g_1(z)$; they arise from
parity violation in electron or quark coupling with
$Z$. The novel spin effects appear in the third
class of terms, which involves the
product of the transversity distributions ${\hat h}_1(z) {\hat h}_1(\bar z)$.
Let us discuss the physics associated with this term
in some detail.

The origin of the $\sin^2\Theta \cos(\varphi+\bar{\varphi}) \,
\hat{h}_1(z) \hat{h}_1(\bar z)$ term is seen
in the expression for the cross
section into $q \bar{q}$, Eq.~(\ref{res1}), in
the term proportional to ($\sigma_x^q\otimes
\sigma^{\bar{q}}_{\bar x}+\sigma^{q}_y\otimes
\sigma^{\bar q}_{\bar y}$).
To appreciate the structure of that particular form,
consider the annihilation through the pure photon
channel. The intermediate state photon is produced with helicity
of $\pm1$, with equal probability for unpolarized incident leptons. If
the $q \bar{q}$ pair were massive quarks produced near threshold, then
their spins would be aligned with the photon spin direction, \ie the beam
direction. At production angle $\Theta$ near $\pi/2$, the term
being considered must be maximum in magnitude. That requires the
$q$ and $\bar{q}$ spins to be preferentially parallel to one another and
to align with the $e^{-}$ or $e^{+}$ beam directions. Recall that the $q$
and $\bar{q}$ momenta are in the $\pm\hat{z}$ direction and the
$\hat{x}$ axis is in the scattering plane. Given the overall
sign of the $Q_q^2$ term, it is seen that the cross section favors the
$\sigma^q_x \otimes \sigma^{\bar q}_{\bar x}$ expectation value be positive.
Since the anti-particle's $\bar x$-axis is oriented parallel to the particle's
$x$-axis, this means that the $x$-components of $q$ and $ \bar{q}$ spin
tend to be aligned. The corresponding $y$-components tend to be
anti-aligned, since the anti-quark's $\bar y$-axis is antiparallel to the
quark's $y$-axis in our convention.

   If the quark spin has an azimuthal orientation specified by $\phi_q$
and the anti-quark, $\bar{\phi}_{\bar{q}}$, then the alignment of spins
just specified leads to cos($\phi_q+\bar{\phi}_{\bar{q}}$) being
positive, which favors the argument near $0$. This result for the
transverse components of the $q \bar{q}$ pair, can be visualized by
having the $\bar{q}$ spin vector reflected through the scattering plane.
Then the transverse spin vector of the quark tends to be parallel to the
transverse spin vector of the {\it reflected\/} $\bar{q}$, \ie the quark
transverse spin tends to line up with the mirror reflected anti-quark
transverse spin.

Next suppose that the quark (anti-quark) spin orientation
is passed on to the $\Lambda$ ($\bar \Lambda$) fragmentation product. The
asymmetry of the $\Lambda$ decay into $p \pi^{-}$ provides a measure of
the spin orientation. As Eq.~(\ref{gary1})
shows, the proton momentum tends to
be aligned with the $\Lambda$ spin.
The corresponding Eq.~(\ref{gary2}) for the
$\bar{\Lambda}$ decay into an antiproton yields the opposite
distribution---the antiproton momentum tends to be antiparallel to the
$\bar{\Lambda}$ spin orientation. Hence, while the $x$-components
($y$-components) of the quark-antiquark spins tend to align (anti-align),
the $x$-components of the decay proton and antiproton momenta tend to
anti-align (align). This is the interpretation of the $\cos(\varphi
+\bar{\varphi})$ term in the cross section of Eq.~(\ref{fin}).

Note that for the $Z^0$ intermediate state the
conclusion is opposite from the above discussion,
because $v_q^2-a_q^2$ is negative. So the transverse momentum
of the proton tends to be aligned with that of
the reflected antiproton.

  This construction leads to a simple phenomenological procedure for
determining the value (for fixed $z$ and $\bar{z}$) of the product
$\hat{h}_1(\bar{z}) \hat{h}_1(z)$. For the photon case, the
above discussion is summarized by the statement that
${\vec{P_p}}{\!}^\bot$ and
${\vec{P_{\bar p}}}{\!}^\bot$
tend to be on the same side of
the scattering plane. For the $Z^0$ case the tendency is for opposite
sides of the scattering plane. So it is natural to define an asymmetry
(for fixed $z$ and $\bar{z}$) via the
number of proton-antiproton pairs on the same side of the scattering
plane minus the number on opposite sides of the plane. The asymmetry
selects the desired term, and has a simple $\Theta$ dependence, so that
results from all $\Theta$ can be combined. The precise expression for
this asymmetry follows from Eq.~(\ref{fin}).

\section{Conclusion}

In this paper, we have obtained
a differential cross section for the process
$e^-e^+\to q\bar q\to\Lambda X\bar\Lambda\bar X\to
p\pi^-X\bar p\pi^+\bar X$. The cross section
depends on, and therefore allows us to extract,
the three twist-two quark fragmentation
functions to the spin-1/2 hyperon, $\hat f_1( z)$,
$\hat g_1( z)$, and $\hat h_1(z)$. These fragmentation
functions contain important information about
the soft QCD physics in quark fragmentation.
Particularly interesting are the spin-dependent
fragmentation functions which encode the behavior
of the spin transfer.
The cross section was obtained through a $2\times 2$
spin density formalism in the helicity basis, in which
the physics is made crystal clear.
The formalism is general and
can be used for other similar spin processes.

The physical process that we discussed is accessible
currently at LEP and SLAC. While up and down quark jets
can produce $\Lambda$ and $\bar \Lambda$ abundantly,
the polarized hyperons are mostly produced
from $s\bar s$ jets, as we expect from the constituent quark model.
The production rate for $s\bar s$ jets is the same as
$d\bar d$ and $b\bar b$. The spin transfer from polarized
$s$ to $\Lambda$ is expected to be large, especially
in the large $z$ region. The weak decay of $\Lambda$ ($\bar \Lambda$)
into a proton and charged pion can be reconstructed easily
and provides an excellent polarization analyzer.
To obtain $\hat h_1(z) \hat h_1(\bar z)$
with enhanced statistics, an asymmetry can be defined by
summing over all events with different $\Omega$ and $P_\perp$'s
in bins of $z$ and $\bar z$. Given these remarks,
we are looking forward to
a first measurement of the spin-dependent fragmentation functions!

\begin{figure}
\caption{Feynman diagram for the process
$e^-e^+\to q\bar q\to\Lambda X\bar\Lambda\bar X\to
p\pi^-X\bar p\pi^+\bar X$. The arrows denote directions of
momentum flow (the fermion number flow for antiparticles
is against the momentum flow). The helicity indices
are explained in the text.}
\label{fig1}

\end{figure}

\begin{references}
\bibitem{AND}
B. Andersson, G. Gustafson, G. Ingleman, and T. Sjostrand,
Phys. Reports. 97 (1983) 31.

\bibitem{JAF}
R.~L.~Jaffe and X.~Ji, Phys. Rev. Lett. {\bf 71}, 2547
(1993).

\bibitem{JI}
X.~Ji, Phys. Rev. {\bf D49}, 114 (1994).

\bibitem{COLLINS}
J.~Collins, Nucl. Phys. B396 (1993) 161.

\bibitem{MUTA}
F. J. Yndurain, {\sl Quantum Chromodynamics}, Springer-Verlag,
New York, 1983.

\bibitem{MUELLER}
A. Mueller, Phys. Rep. 73C (1981) 237.

\bibitem{JW}
M. Jacob and G. C. Wick, Ann. Phys. 7 (1959) 404.

\bibitem{BD}
J.~Bjorken and Drell, {\sl Relativistic Quantum Fields}
(McGraw-Hill, New York, 1965).

\bibitem{CB}
E.~D.~Commins AND P.~H.~Bucksbaum, {\sl Weak interactions
of leptons and quarks}, (Cambridge University Press, 1983).

\bibitem{PDG}
Particle Data Group, Phys. Rev. D50 (1994) 1173.
\end{references}
\end{document}